\documentclass{ws-p8-50x6-00}

\begin{document}

\title{Strangeness Production as a Diagnostic Tool for Understanding Heavy Ion Reactions}

\author{J. L. Nagle}

\address{Nevis Laboratories, Columbia University, New York, NY 10027, USA\\ 
E-mail: nagle@nevis.columbia.edu}


\maketitle

\abstracts{
Strangeness production has long been proposed as a diagnostic tool 
for understanding the dynamics of relativistic heavy ion collisions.
In this presentation we review the traditional picture of strangeness 
enhancement as a signature for quark-gluon plasma formation.  We then review,
in order, some experimental data on strange particle production in 
$e^{+}e^{-}$, $pp$, $p\overline{p}$, proton-nucleus and nucleus-nucleus 
collisions.  This is not
a comprehensive review, but rather an emphasis of a few
significant points.  Any clear interpretation of strange particle yields measured
in heavy ion reactions is impossible without a physical understanding of the production
mechanisms in elementary particle collisions.   
}

\section{Introduction}

Hadronization of parton jets and multi-partonic systems is not directly calculable using 
Quantum Chromodynamics (QCD).  This non-perturbative physics can only currently be described
with phenomenological models that have been developed over the last 30 years.  There are
many such models including string fragmentation models, quark coalescence models, and statistical models.
For the purposes of this presentation, we will use the statistical model as a baseline for 
comparing different colliding systems in terms of strangeness production.  This is not meant
to advocate one schematic description of hadronization over another.

\begin{figure}[t]
%\figurebox{20pc}{15pc}{} % to have a box alone
\epsfxsize=20pc % will enlarge or reduce the postscript figures based on the xsize
\epsfbox{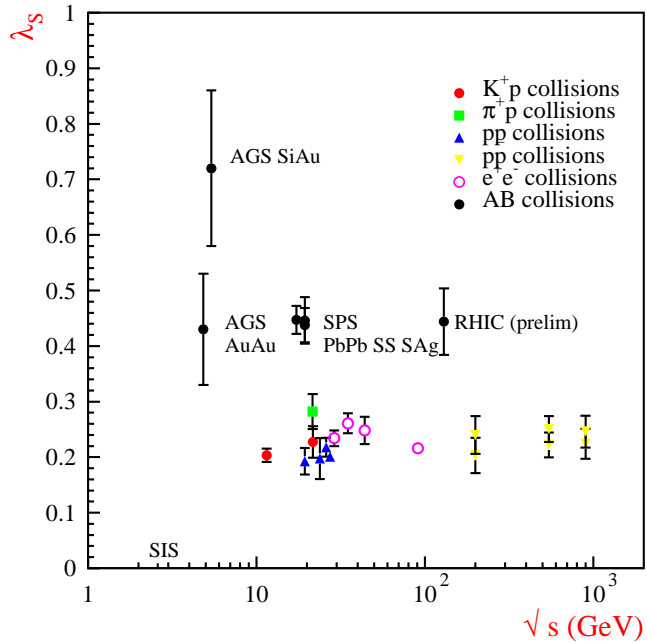} % postscript image file name
\caption{$\lambda_{s}$ as a function of $\sqrt{s}$ for various colliding systems\label{fig:lam}}
\end{figure}

Calculations assuming thermal and chemical equilibrium in hadron gas fireballs 
assume a universal hadronization
mechanism and find a parton-hadron transition at a 
temperature of $T \approx 170$ MeV.\cite{Becattini:1996gy}
The agreement between experimentally observed hadron yields in electron-positron
jet fragmentation and in proton-(anti)proton reactions is reasonable.  However, it is observed that
particles carrying strange quarks are not in complete chemical equilibrium.  There appears 
to be an additional suppression as expressed below:
\begin{equation}
\lambda_{s} = {{s\overline{s}} \over {{{1}\over{2}}{(u\overline{u} + d\overline{d})}}}
\end{equation}  
There appears to be a deficit of strange hadrons that is common
in inclusive proton-proton, proton-antiproton, $e^{+}e^{-}$ collisions as shown in
Figure~\ref{fig:lam}.

It was this observed suppression that led to the postulation 
that one could bring the strange hadrons into chemical equilibrium 
via pre-hadronization partonic interactions or post-hadronization 
re-scattering.\cite{Koch:1986ud}  
However, the time scale for
total strangeness ($\Lambda$, $K$) equilibration via hadronic re-scattering 
is 10-100 fm/c, while for strange antibaryons 
($\overline{\Lambda}, \overline{\Xi}, \overline{\Omega}$) 
take more than 1000 fm/c.  It is interesting to note that some recent calculations
differ substantially from the above rate calculations.\cite{Greiner:2000vb} 
  
The strangeness equilibration time in a deconfined system of a quark-gluon plasma is substantially
shorter, of order 3-6 fm/c.  This is due to contributions from gluon-gluon fusion to $s\overline{s}$
pairs.  This time is well within the expected heavy ion collision lifetime of 
10-15 fm/c before particles are free streaming.  
It is comparable to the time in heavy ion collisions that 
one might naively expect a deconfined region to exist before
it cools to the point of hadronization.

\section{Proton-Proton}
Often the assumption is made that there is no quark-gluon plasma formed in proton induced
reactions.  This raises the question of the definition of the quark-gluon plasma.
One answer is that it is an extended region of space with strong color fields characterized
by a large density of deconfined partons and restoration of approximate chiral symmetry.
Although the parton density achieved in elementary particle collisions is quite large, the 
volume is not, and one can argue whether it represents a well-defined thermodynamic
state.

In fact, Bjorken speculated that in the 
``interiors of large fireballs produced in 
very high-energy $p\overline{p}$ 
collisions, vacuum states of the strong interaction are 
produced with anomalous chiral order parameters.''\cite{Bjorken:1993cz}  
He applied a schematic picture referred to as the Baked Alaska in 
describing $p\overline{p}$ collisions, and the MINIMAX experiment at Fermilab 
searched for disoriented chiral condensates.\cite{Brooks:2000xy}  
Although MINIMAX yielded a null result, this picture is not ruled out.

Experiment E735 at the Fermilab Tevatron studied $p\overline{p}$ collisions to look for 
quark-gluon plasma signatures.   Both E735 and UA1 and others observed a substantially 
larger source volume in high multiplicity $pp$ ($p\overline{p}$) events via two particle 
correlations measurements.  They also observed transverse momentum
spectra that have a mass ordering that some described via hydrodynamic flow. 
E735, UA1 and the AFS experiment at the ISR observed a substantial enhancement 
of strange particle production as a function
of the total particle multiplicity.  These observations led some to speculate that
a quark-gluon plasma was being formed.\cite{Levai:1992ku}  
However, these effects can also be explained consistently 
by a picture of event bias from the multiplicity selection on events with hard processes,
in particular gluon jets.\cite{Wang:1992us}  
Thus, one needs to take care in comparing non-inclusive $pp$ data.

It is also interesting that statistical models often use the grand canonical ensemble.  
One can use the GCE even when the energy and
other quantum numbers are conserved.  The temperature and chemical potentials simply reflect 
the characteristics of the system.  However one must be careful in the interpretation 
in particular of fluctuations quantities.  
If the volume of the system is large GCE is appropriate; however, for small volumes
you must conserve quantum numbers (for example strangeness) in every event.  
Thus the canonical ensemble is relevant.  In the CE, strangeness is suppressed in 
very small volume reactions and reaches the GCE limit only for
large volume.  It would be interesting to see if this can be tested in 
$pp$ events with different HBT radii.  There may be technical difficulties in
dealing with auto-correlations with HBT radii and other collision characteristics.

\section{Proton-Nucleus}

Inclusive $pA$ reactions have only a few binary $NN$ collisions, and are thus 
not so different than $pp$.  However, measuring recoil nucleons in $pA$ reactions allows
one to select events where the probability of
many binary collisions is large.\cite{Chemakin:1999jd}
This type of analysis is now being done by experiment E910 at the
BNL-AGS and NA49 at the CERN-SPS.  

\begin{figure}[t]
%\figurebox{20pc}{15pc}{} % to have a box alone
\epsfxsize=15pc % will enlarge or reduce the postscript figures based on the xsize
\epsfbox{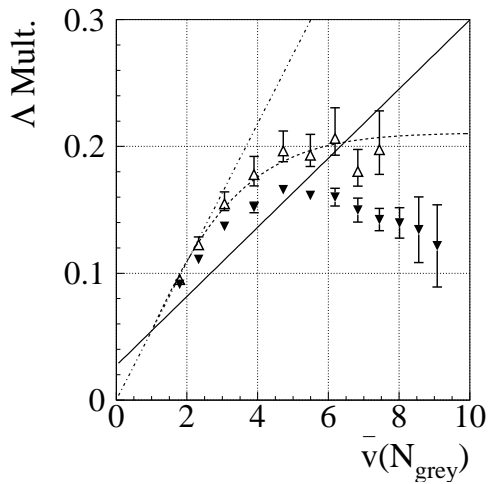} % postscript image file name
\caption{Measured $\Lambda$ multiplicity as a function of the average number
of binary collisions $\nu$\label{fig:910}}
\end{figure}

E910 observes that the $K^{+}$ and $\Lambda$ production increases rapidly as a function
of the number of collisions suffered by the proton and then appears to level off with 
no increase in production after $\approx4$ collisions as 
shown in Figure~\ref{fig:910}.\cite{Chemakin:2000ha}.  
The data do not appear
to scale with the number of participating nucleons (as is often used as
a baseline for comparison in AA collisions).  E910 has pointed out
that the difference between the $\Lambda$ production data and participant scaling
appears to increase linearly with the number of binary collisions up to three collisions.
Three step valence quark stripping mechanisms and baryon junction models
have been suggested to explain the data.
%\begin{equation}
%N_{\Lambda}^{proj} = N_{coll} \times {1 \over 2} N_{\Lambda}^{pp} (\rm{for~N_{coll} \le 3})	
%\end{equation}
The $\Xi$ is also measured and shows an even stronger enhancement 
relative to participant scaling that the
singly strange hadrons.  Also, most of the additional production is at target
rapidity.  Thus, measurements only at mid-rapidity are missing most of the increased production.
$K^{-}$ production increases with $N_{coll}$, but after 4-5 collisions it sharply
decreases.  This decrease is potentially explained in that when the proton strikes the 
center of the nucleus ($N_{coll}>5$), there is a high probability for
processes such as $K^{-} + N \longrightarrow \pi + \Lambda$ to effectively absorb the $K^{-}$.

NA49 also observes a similar enhancement for all strange particles, in particular the
the doubly strange baryons.  NA49 and WA97 observe that the $\Xi$ yields increase 
faster than would be expected from simple participant scaling.
Understanding the physical mechanism for
the strangeness enhancement at both AGS and SPS energies is a theoretical challenge.

\section{Nucleus-Nucleus}

Heavy ion collisions at the AGS show an increase in the $K^{+}/\pi^{+}$ 
ratio as a function of centrality or the
number of participating nucleons, and thus total strangeness is enhanced.  
Most hadron yields can be
described by a statistical equilibrium model within systematic errors, and one can see in
Figure~\ref{fig:lam} the value for $\lambda_{s}$.
At the lowest energies run at the AGS ($p_{beam}=2~GeV$), strangeness production
is near threshold and the yield drops drastically as expected.

Experiment E917 measures a ratio 
${{\overline{\Lambda}} \over {\overline{p}}} = 3.6 ^{+4.7}_{-1.8}$,
which confirms an earlier indirect measure by experiments E864 and E878.  This large
enhancement of strange antibaryons is above the statistical equilibrium level and may 
result from some balance of enhanced production of $\overline{\Lambda}$ and 
annihilation losses for the $\overline{p}$.  Unfortunately there is no measurement 
of $\overline{\Xi}$.  This is an excellent possible future measurement at the Japanese
Hadron Facility (JHF) or an energy upgraded GSI.

Many experiments at the CERN-SPS also observe enhancement of strangeness in many channels.
In particular, WA97 and NA49 observe a very large increase in the production
of $\Xi$, $\Omega$, and $\overline{\Xi}$, $\overline{\Omega}$.
The enhancement of participant scaling increases with increasing strangeness content
$E_{\Lambda} < E_{\Xi} < E_{\Omega}$.  This is the same trend seen in $pA$.
The statistical models give good agreement with a temperature of 170 MeV, in agreement
with that obtained from high energy $pp$, $p\overline{p}$ and $e^{+}e^{-}$.

The press release from CERN on the discovery of the quark-gluon
plasma states that, ``Since the hadron abundances appear 
to be frozen in at the point of hadron formation, this enhancement 
signals a new and faster strangeness-producing process before or 
during hadronization, involving intense re-scattering 
among quarks and gluons.''  It seems that this conclusion is made without full
consideration of data at lower energy and in $pA$ reactions.

At RHIC there is already preliminary data from all four experiments on strange
and non-strange hadron yields.  They also give a reasonable fit by 
a statistical model assuming equilibration of strangeness with T = 170 MeV.  
The enhancement of strangeness appears remarkably similar 
at the AGS, CERN-SPS and RHIC.  I predict that LHC will see similar results.

\section{Conclusions}

Almost everywhere we look we observe enhancement of strange particle production.  
This effect is magnified as we go from proton-proton to proton-nucleus to 
nucleus-nucleus collisions.  
Many in the field assume this implies that strangeness production cannot be a signal of 
quark-gluon plasma formation.   I believe this conclusion is premature.
We must understand the global hadronization process from a region of vacuum with strong 
color fields in order to make progress.  Future data at RHIC in $pp$, $pA$ and $AA$
are eagerly awaited.

\section*{Acknowledgments}
Useful discussions with Ken Barish, Sean Kelly, Dave Morrison and X.N. Wang are acknowledged.

\end{document}